\documentclass[a4,11pt]{article} 
\usepackage{amssymb}
\usepackage{amsmath}
\usepackage[pdftex]{graphicx}
\usepackage{cite}
\usepackage{slashed}
\usepackage{bm}
\usepackage{fancyhdr}
\usepackage{float}
\usepackage{tocloft}
\usepackage[top=2.5cm, bottom=2.0cm, left=2.5cm, right=2cm]{geometry}

\begin{document}

\begin{center}
{\large \bf	The influence of the couplings on the Z-production at ILC in the Randall - Sundrum model}\\

\vspace*{1cm}

 { Bui Thi Ha Giang$^{a,}$  \footnote{giangbth@hnue.edu.vn}}\\

\vspace*{0.5cm}
 $^a$ Hanoi National University of Education, 136 Xuan Thuy, Hanoi, Vietnam
\end{center}

\begin{abstract}
An attempt is made to present the influence of the couplings on $e^{-}e^{+}$  scattering process with polarized initial beams in the Randall-Sundrum (RS) model. We have studied the contribution of radion, Higgs and gauge bosons on the Z-production cross-section at International Linear Colliders (ILC). The cross-section depends strongly on the polarization of $e^{-}$, $e^{+}$  initial beams, the center of mass energy $\sqrt{s} $ and model interaction parameters. The results indicate that with the influence of the couplings, including radion and Higgs in RS model, the cross-sections for the Z production are enhanced, in which the u, t-channels have given  dominated contributions.  
\end{abstract}

\section{Introduction}
\hspace*{1cm} The Standard model (SM) has been quite successful in elementary particle physics. The discovery of the Higgs boson at the Large Hadron Collider (LHC) has completed the Standard model \cite{gaa, cha}.  However, the existence of some theoretical drawbacks has motivated the models beyond the SM. Most of the beyond the SM have expected either new particles or new couplings. The Randall-Sundrum model has been one of the extended SM models. The RS model involved two three-branes has allowed the existence of an additional scalar called the radion ($\phi $ ) \cite{rs}. Based on the same quantum numbers, radion and Higgs boson can be mixed \cite{jung, eboos, frank, ali}. The radion couplings to $\gamma \gamma ,gg$ and $Z\gamma $  are dominated in the region $\xi \,\left[0,0.3\right]$ \cite{ahm}. It is of particular interest to analyse the radion couplings in conformal limit $\xi = 1/6$.\\
\hspace*{1cm} The trilinear gauge boson couplings have been important in testing the electroweak interactions \cite{atag}. Diboson production, in particular ZZ and $W^{+}W^{-}$, are also extensively used in Higgs boson measurements \cite{kall}. Moreover, the anomalous vertices, including ZZZ, $\gamma ZZ$, $\gamma \gamma Z$ interactions, which are not present at tree level in SM, have been widely discussed in the different colliders: $e^{-}e^{+}$ collider, $\gamma e^{-}$ collider, hadron collider \cite{atag, goun, bau, anan1, anan2, anan3, raha, cas}. In experiment, the cross-section for Z production in $p\bar{p}$ collisions has been measured by both the ATLAS and CMS collaboration \cite{abd, 7atlas, 7cms, 13atlas, 8atlas}. Because of clean electron and positron sources at ILC, Z boson produced at the high energy collisions could give the possible measurement. Any possible new physics in the Z boson production collision is expected to change the cross-section. \\
\hspace*{1cm} In our present work, we have studied $e^{-}e^{+} \rightarrow ZZ \rightarrow l^{+} l^{-} q\overline{q}$, $e^{-}e^{+} \rightarrow \gamma Z \rightarrow \gamma l^{+} l^{-} \left(\gamma q\bar{q}\right)$  processes, including the vertices of Z boson as ZZZ, $\gamma ZZ$, $\gamma\gamma Z$, $\phi ZZ$, $hZZ$, $\gamma Zh$, $\gamma Z \phi$. With the contribution of new anomalous interaction in RS model, including radion and Higgs, the total cross-section has been expected to enhance. The layout of this paper is as follows. \textbf{In Section II, we review the Randall-Sundrum model and the mixing of Higgs - radion.} The influence of the new couplings on the Z production at ILC is calculated in Section III. Finally, we summarize our results and make conclusions in Section IV.
\section{A review of Randall-Sundrum model and the mixing of Higgs-radion}
\hspace*{1cm}The RS model consists of one extra dimension bounded by two 3-branes, this can be viewed as an $S^{1} /Z_{2}$ orbifold \cite{ahm}. The fifth dimension is bounded at UV-brane located at $y = 0$ and the IR-brane located at $y = \pi r_{c}$. The five dimensional metric has the following form:
\begin{equation}
ds^{2} = e^{-2kb_{0}|y|}\eta_{\mu\nu}dx^{\mu}dx^{\nu} - b_{0}^{2}dy^{2},
\end{equation}
where $b_{0}$ is a length parameter for the 5th dimension, $r_{c}$ is the compactification radius and $k$ is the curvature of the 5D geometry. The exponential represents the warp factor which generates the gauge hierarchy. The values of the bare parameters are determined by the Planck scale and the applicable value for size of the extra dimension is assessed by $kr_{c} \simeq 11 - 12$. Thus the weak and the gravity scales can be naturally generated. The relation $\overline{M}_{Pl}^{2} = M_{5}^{3}/k$ is derived from the 5D action. The scale of physical phenomena on the IR-brane is given by $\Lambda_{\phi} \equiv \overline{M}_{Pl} e^{-kr_{c}\pi}$ with $\Lambda _{\phi } \sim $ few TeV. The mass of the $n^{th}$ KK graviton excitation is $m^{G}_{n} = x_{n}^{G}k\Lambda_{\phi}/\overline{M}_{Pl}$, with $x_{n}^{G}$ being the roots of the $J_{1}$ Bessel function \cite{rue}. The coupling strength of the graviton KK states to the SM fields is $\Lambda_{\phi}^{-1}$.
Gravitational fluctuations around the background metric will be defined by replacing:
\begin{equation}
\eta_{\mu\nu} \rightarrow \eta_{\mu\nu} + \epsilon h_{\mu\nu} (x, y) \qquad     b_{0} \rightarrow b_{0} + b(x).
\end{equation}
The gravity-matter interactions have been obtained \cite{kow}
\begin{equation}
\mathcal{L}_{int} = -\dfrac{1}{\hat{\Lambda}_{W}}\Sigma_{n\neq 0} h^{n}_{\mu\nu} T^{\mu\nu} - \dfrac{\phi_{0}}{\Lambda_{\phi}} T^{\mu}_{\mu},
\end{equation}
where $h^{n}_{\mu\nu} (x)$ are the Kaluza-Klein (KK) modes of the graviton field $h_{\mu \nu}(x, y)$, $\phi_{0} (x)$ is the radion field, $\Lambda_{\phi} \equiv \sqrt{6}M_{Pl}\Omega_{0}$ is the VEV of the radion field and $\hat{\Lambda}_{W} \equiv \sqrt{2}M_{Pl}\Omega_{0}$. The $T^{\mu\nu}$ is the energy-momentum tensor, which is given at the tree level \cite{bae, csa}
\begin{equation}
T^{\mu}_{\mu}=\Sigma_{f} m_{f} \overline{f}f - 2m^{2}_{W}W^{+}_{\mu}W^{-\mu}-m^{2}_{Z}Z_{\mu}Z^{\mu} + (2m^{2}_{h_{0}}h_{0}^{2} - \partial_{\mu}h_{0}\partial^{\mu}h_{0}) + ...
\end{equation}
The gravity-scalar mixing is described by the following action\cite{domi}
\begin{equation}
S_{\xi } =\xi \int d^{4}x \sqrt{g_{vis} } R(g_{vis} )\hat{H}^{+} \hat{H},
\end{equation}
where $\xi $ is the mixing parameter, $R(g_{vis})$ is the Ricci scalar for the metric $g_{vis}^{\mu \nu } =\Omega _{b}^{2} (x)(\eta ^{\mu \nu } +\varepsilon h^{\mu \nu } )$ induced on the visible brane, $\Omega _{b} (x) = e^{-kr_{c} \pi} (1 + \frac{\phi_{0}}{\Lambda _{\phi }})$ is the warp factor, $\hat{H}$ is the Higgs field in the 5D context before rescaling to canonical normalization on the brane.
 With $\xi \ne 0$, there is neither a pure Higgs boson nor pure radion mass eigenstate. This $\xi$ term mixes the $h_{0}$ and $\phi_{0}$ into the mass eigenstates $h$ and $\phi$ as given by 
\begin{equation} 
\left(\begin{array}{c} {h_{0} } \\ {\phi _{0} } \end{array}\right)=\left(\begin{array}
{cc} {1} & {6\xi \gamma /Z} \\ {0} & {-1/Z} \end{array}\right)\left(\begin{array}{cc}
 {\cos \theta } & {\sin \theta } \\ {-\sin \theta } & {\cos \theta } \end{array}\right)
 \left(\begin{array}{c} {h} \\ {\phi } \end{array}\right)=\left(\begin{array}{cc}
  {d} & {c} \\ {b} & {a} \end{array}\right)\left(\begin{array}{c} {h} \\ {\phi } \end{array}\right), \label{pt}
\end{equation}
where
$Z^{2} = 1 + 6\gamma ^{2} \xi \left(1 -\, \, 6\xi \right) = \beta - 36\xi ^{2}\gamma ^{2}$ is the coefficient of the radion kinetic term after undoing the kinetic mixing, $\gamma = \upsilon /\Lambda _{\phi }, \upsilon = 246$ GeV, $a = -\dfrac{cos\theta}{Z}, b = \dfrac{sin\theta}{Z}, c = sin\theta + \dfrac{6\xi\gamma}{Z}cos\theta, d = cos\theta - \dfrac{6\xi\gamma}{Z}sin\theta$. The mixing angle $\theta $ is
\begin{equation}
\tan 2{\theta } = 12{\gamma \xi Z}\frac{m_{h_{0}}^{2}}{m_{\phi _{0}}^{2} - m_{h_{0}}^{2} \left( Z^{2} - 36\xi^{2} \gamma ^{2} \right)},
\end{equation}
where $m_{h_{0}}$ and $m_{\phi _{0}}$ are the Higgs and radion masses before mixing.\\
The new physical fields h and $\phi $ in (\ref{pt}) are Higgs-dominated state and radion, respectively
\begin{equation} 
m_{h,\phi }^{2} =\frac{1}{2Z^{2} } \left[m_{\phi _{0} }^{2} +\beta m_{h_{0} }^{2} \pm \sqrt{(m_{\phi _{0} }^{2} +\beta m_{h_{0} }^{2} )^{2} -4Z^{2} m_{\phi _{0} }^{2} m_{h_{0} }^{2} } \right].
\end{equation}
\\
 There are four independent parameters $\Lambda _{\phi } ,\, \, m_{h} ,\, \, m_{\phi } ,\, \, \xi$ that must be specified to fix the state mixing parameters. We consider the case of $\Lambda _{\phi } = 5$ TeV and $\frac{m_{0} }{M_{P} } = 0.1$, which makes the radion stabilization model most natural \cite{csa}.
\section{The influence of the couplings on the Z production at ILC}
\hspace*{1cm} Vector boson pair production is interesting in itself, including tri-vector boson couplings \cite{tom}. An investigation of diboson production at ILC plays an important role in testing the SM and searching for physics beyond. In our previour work \cite{mpla}, we have evaluated the contribution of scalar unparticle on the W-pair production in the RS model. In this work, we will evaluate the significance of the new couplings on the Z production, including the $e^{-}e^{+} \rightarrow ZZ \rightarrow l^{+} l^{-} q\bar{q}$ and $e^{-}e^{+} \rightarrow \gamma Z \rightarrow \gamma l^{+} l^{-} \left(\gamma q\bar{q}\right)$ processes with the polarized initial beams at ILC. An $e^{+}e^{-}$ collider is uniquely capable of operation at series of energies near the threshold of a new physics process. This is an extremely powerful tool for precision measurements of particle masses and unambiguous particle spin determinations \cite{kiku1}.\\
\subsection{The $e^{-}e^{+} \rightarrow ZZ \rightarrow l^{+} l^{-} q\bar{q}$ collision}
\begin{figure}[!htb] 
\begin{center}
\includegraphics[width= 14 cm,height= 5 cm]{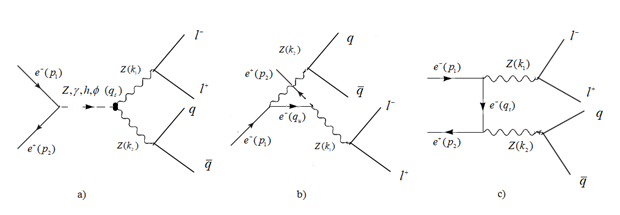}
\caption{\label{feynmanZZ} Feynman diagrams for $e^{+}e^{-} \rightarrow ZZ \rightarrow l^{-}l^{+}q \overline{q}$ collision,representing the s, u, t-channels, respectively.}
\end{center}
\end{figure}

\hspace*{1cm}We consider the collision process in which the initial state contains electron and positron, the final state contains a pair of Z boson  
\begin{equation} \label{pt1}
e^{-}(p_{1}) + e^{+}(p_{2}) \    \rightarrow        \  Z (k_{1}) + Z (k_{2}).
\end{equation}
Here, $p_{i}, k_{i}$ (i = 1, 2) stand for the momentums. There are three Feynman diagrams contributing to reaction (\ref{pt1}), representing the s, u, t-channels exchange depicted in Fig.\ref{feynmanZZ}.\\
\hspace*{1cm}The transition amplitude representing the s-channel is given by
\begin{equation}
M_{s} = M_{Z} + M_{\gamma} + M_{h} + M_{\phi},
\end{equation}
here 
\begin{align}
&M_{Z} = \dfrac{-\overline{g}_{eZ}}{q^{2}_{s} - m^{2}_{Z}} \varepsilon^{*}_{\mu} (k_{1}) \Gamma^{\sigma\mu\nu}_{ZZZ}(q_{s}k_{1} k_{2})\varepsilon^{*}_{\nu} (k_{2})\left(\eta_{\sigma\beta} - \dfrac{q_{s\sigma}q_{s\beta}}{m^{2}_{Z}}\right)\overline{v}(p_{2}) \gamma^{\beta}\left(v_{e} - a_{e}\gamma^{5}\right) u(p_{1}),\\
&M_{\gamma} = \dfrac{-e}{q^{2}_{s}} \varepsilon^{*}_{\mu} (k_{1}) \Gamma^{\sigma\mu\nu}_{\gamma ZZ}(q_{s}k_{1}k_{2})\varepsilon^{*}_{\nu} (k_{2})\eta_{\sigma\beta} \overline{v}(p_{2}) \gamma^{\beta} u(p_{1}),\\
&M_{h} = \dfrac{\overline{g}_{eeh}\overline{g}_{hZ}}{q^{2}_{s} - m^{2}_{h}}\varepsilon^{*}_{\mu} (k_{1}) \left[\eta^{\mu\nu} - 2g^{Z}_{h}\left(k_{1}k_{2}\eta^{\mu\nu} - k_{1}^{\nu}k_{2}^{\mu}\right)\right]\varepsilon^{*}_{\nu} (k_{2})\overline{v}(p_{2})u(p_{1}),\\
&M_{\phi} = \dfrac{\overline{g}_{ee\phi}\overline{g}_{\phi Z}}{q^{2}_{s} - m^{2}_{\phi}}\varepsilon^{*}_{\mu} (k_{1}) \left[\eta^{\mu\nu} - 2g^{Z}_{\phi}\left(k_{1}k_{2}\eta^{\mu\nu} - k_{1}^{\nu}k_{2}^{\mu}\right)\right]\varepsilon^{*}_{\nu} (k_{2})\overline{v}(p_{2})u(p_{1}).
\end{align}
\hspace*{1cm}The transition amplitude representing the u-channel can be written as
\begin{equation}
M_{u} = -i\dfrac{|\overline{g}_{eZ}|^{2}}{{q^{2}_{u} - m^{2}_{e}}}\overline{v}(p_{2})\gamma^{\mu}\left(v_{e}-a_{e}\gamma^{5}\right)\varepsilon^{*}_{\mu} (k_{1})\left(\slashed{q}_{u}+m_{e}\right)\gamma^{\nu}\left(v_{e}-a_{e}\gamma^{5}\right) \varepsilon^{*}_{\nu} (k_{2})u(p_{1}).
\end{equation}
\hspace*{1cm}The transition amplitude representing the t-channel is given by
\begin{equation}
M_{t} = -i\dfrac{|\overline{g}_{eZ}|^{2}}{{q^{2}_{t} - m^{2}_{e}}}\overline{v}(p_{2})\gamma^{\nu}\left(v_{e}-a_{e}\gamma^{5}\right)\varepsilon^{*}_{\nu} (k_{2})\left(\slashed{q}_{t}+m_{e}\right)\gamma^{\mu}\left(v_{e}-a_{e}\gamma^{5}\right) \varepsilon^{*}_{\mu} (k_{1})u(p_{1}).
\end{equation}
Here, $\overline{g}_{eZ}, \overline{g}_{hZ}, \overline{g}_{\phi Z}$ can be found in \cite{ahm, domi, bgrz}.\\
Feynman rules for the couplings in the RS model are showed as follows \cite{ahm}
\begin{align}
\label{eeh}&\overline{g}_{eeh} = -\dfrac{gm_{e}}{2m_{W}}\left( d + \gamma b\right),\\
\label{eep}&\overline{g}_{ee\phi} = -\dfrac{gm_{e}}{2m_{W}}\left( c + \gamma a\right),\\
&g_{hZZ}^{\mu \nu} = i\overline{g}_{hZ}\left[\eta^{\mu\nu} - 2g^{Z}_{h}\left(k_{1}k_{2}\eta^{\mu\nu} - k_{1}^{\nu}k_{2}^{\mu}\right)\right],\\
&g_{\phi ZZ}^{\mu \nu} = i\overline{g}_{\phi Z}\left[\eta^{\mu\nu} - 2g^{Z}_{\phi}\left(k_{1}k_{2}\eta^{\mu\nu} - k_{1}^{\nu}k_{2}^{\mu}\right)\right],
\end{align}
where a, b, c, d are the state mixing parameters.\\
The triple gauge boson couplings are given by \cite{raha}
\begin{equation} \label{pt2}
\begin{aligned}
\Gamma^{\sigma\mu\nu}_{\gamma ZZ}&(q_{s}k_{1}k_{2}) =\\
& \dfrac{g_{e}}{m_{Z}^{2}}\Biggl[f_{4}^{\gamma}\left(k_{2}^{\mu}\eta^{\sigma\nu} + k_{1}^{\nu}\eta^{\sigma\mu}\right) q_{s}^{2} - q_{s}^{\sigma} \left(k_{1}^{\nu}q_{s}^{\mu} + k_{2}^{\mu}q_{s}^{\nu}\right) + f_{5}^{\gamma}\left(q_{s}^{\sigma}q_{s\beta}\varepsilon^{\mu\nu\alpha\beta} + q_{s}^{2}\varepsilon^{\sigma\mu\nu\alpha} \right)(k_{1} - k_{2})_{\alpha} \\
& + h_{1}^{Z} \Biggl(k_{2}^{\sigma}q_{s}^{\mu}k_{2}^{\nu} + k_{1}^{\sigma}k_{1}^{\mu}q_{s}^{\nu} + \left(k_{1}^{2} - k_{2}^{2}\right)\left(q_{s}^{\mu} \eta^{\sigma\nu} - q_{s}^{\nu}\eta^{\sigma\mu}\right)  - k_{2}^{\nu}\eta^{\sigma\mu}(k_{2}q_{s}) - k_{1}^{\mu}\eta^{\sigma\nu}(k_{1}q_{s})\Biggr)\\
& - h_{3}^{Z}\left(k_{1}^{\mu}k_{1\beta}\varepsilon^{\sigma\nu\alpha\beta} + k_{2\beta}k_{2}^{\nu}\varepsilon^{\sigma\mu\alpha\beta} + \left(k_{2}^{2} - k_{1}^{2}\right)\varepsilon^{\sigma\mu\nu\alpha}\right)q_{s\alpha}\Biggr].
\end{aligned}
\end{equation}
\begin{equation} \label{pt3}
\begin{aligned}
\Gamma^{\sigma\mu\nu}_{ZZZ}&(q_{s}k_{1}k_{2}) =\\
& \dfrac{g_{e}}{m_{Z}^{2}}\Biggl[f_{4}^{Z} \Biggl(-q_{s}^{\sigma}q_{s}^{\mu}k_{1}^{\nu} - k_{2}^{\sigma}q_{s}^{\mu}k_{2}^{\nu} - k_{2}^{\sigma}k_{1}^{\mu}k_{1}^{\nu} - k_{1}^{\sigma}k_{2}^{\mu}k_{2}^{\nu} -\left(q_{s}^{\sigma}k_{2}^{\mu} + k_{1}^{\sigma}k_{1}^{\mu} \right)q_{s}^{\nu} + \eta^{\sigma\mu} \left(q_{s}^{2}k_{1}^{\nu} + k_{1}^{2}q_{s}^{\nu}\right)\\
&+\eta^{\sigma\nu}\left(q_{s}^{2}k_{2}^{\mu} + k_{2}^{2}q_{s}^{\mu}\right) + \eta^{\mu\nu}\left(k_{2}^{2}k_{1}^{\sigma} + k_{1}^{2}k_{2}^{\sigma}\right)\Biggr) -f_{5}^{Z}\Biggl(\varepsilon^{\sigma\mu\alpha\beta}\left(k_{1}-q_{s}\right)_{\alpha} k_{2\beta}k_{2}^{\nu} + \varepsilon^{\sigma\mu\nu\alpha} \left(k_{1}^{2} - k_{2}^{2}\right)q_{s\alpha}\\
&+ \left(k_{2}^{2} - q_{s}^{2}\right)k_{1\alpha} + \left(q_{s}^{2} - k_{1}^{2}\right)k_{2\alpha} + k_{1\beta}k_{1}^{\mu}\left(k_{2}-q_{s}\right)_{\alpha}\varepsilon^{\sigma\nu\alpha\beta} + q_{s\beta}q_{s}^{\sigma}\left(k_{2} - k_{1}\right)_{\alpha}\varepsilon^{\mu\nu\alpha\beta}\Biggr)\Biggr].
\end{aligned}
\end{equation}
The total cross-section for the whole process can be calculated as follow \cite{raha}
\begin{equation}
\sigma = \sigma (e^{-} e^{+} \rightarrow  ZZ) \times 2 Br(Z\rightarrow l^{-}l^{+}) Br(Z\rightarrow q \overline{q}),
\end{equation}
where 
\begin{equation}
\frac{d\sigma (e^{-} e^{+} \rightarrow  ZZ)}{d(cos\psi)} = \frac{1}{32 \pi s} \frac{|\overrightarrow{k}_{1}|}{|\overrightarrow{p}_{1}|} |M_{fi}|^{2}
\end{equation}
is the expressions of the differential cross-section \cite{pes}. $\psi = (\widehat{\overrightarrow{p}_{1}, \overrightarrow{k}_{1}})$ is the scattering angle.\\
\hspace*{1cm}For numerical calculations, we choose ILC running at a center-of-mass energy of 500 GeV and luminosity $\mathcal{L} = 100fb^{-1}$ \cite{anan3}. The vacuum expectation value (VEV) of the radion field is $\Lambda_{\phi} = 5$ (TeV) \cite{domi}. The radion mass has been selected $m_{\phi} = 10$ GeV \cite{soa1}. The Higgs mass $m_{h} = 125$ GeV (CMS). We give estimates for the cross-sections as follows \\
\hspace*{1cm}i) In Fig.\ref{Fig.2}, the total cross-section is plotted as the function of $P_{e^{-}}, P_{e^{+}}$, which are the polarization coefficients of $e^{-}, e^{+}$ beams, respectively. \textbf{The maximum value of anomalous couplings in the tightest limits (at $1\sigma$ level) with the corresponding observable are chosen as} $f_{4}^{\gamma} = 2.4 \times 10^{-3}$, $f_{4}^{Z} = 4.2 \times 10^{-3}$, $f_{5}^{\gamma} = 2.7 \times 10^{-3}$, $f_{5}^{Z} = 8.8 \times 10^{-3}$, $h_{1}^{\gamma} = 3.6 \times 10^{-3}$, $h_{3}^{\gamma} = 1.3 \times 10^{-3}$, $h_{1}^{Z} = 2.9 \times 10^{-3}$, $h_{3}^{Z} = 2.8 \times 10^{-3}$ \cite{raha}. The collision energy is chosen as $\sqrt{s} = 500$ GeV (ILC). The figure indicates that the total cross-section achieves the maximum value when $P_{e^{-}} = P_{e^{+}} = \pm 1$ and the minimum value when $P_{e^{-}} = 1, P_{e^{+}} = -1$ or $P_{e^{-}} = -1, P_{e^{+}} = 1$. This result reverses the consequence of $e^{+}e^{-}$ $\rightarrow$ $W^{+}W^{-}$ collision in our previous work. \textbf{The  possible value of the cross section to be 41.323 fb, obtained for $f_{4}^{\gamma, Z} = 0$, $f_{5}^{\gamma} = 2 \times 10^{-4}$, $f_{5}^{Z} = 3.2 \times 10^{-3}$, $h_{1}^{\gamma, Z} = 0$, $h_{3}^{\gamma, Z} = 0$, $\sqrt{s} = 500$ GeV, $P_{e^{-}} = P_{e^{+}} = \pm 1$. This value is larger than the observable theoretical value (38.096 fb) in Ref.\cite{raha} due to the radion and Higgs contributions.}  \\
\hspace*{1cm}ii) \textbf{The couplings of the radion and Higgs to electrons are proportional to the electron mass, as is clear from equations (\ref{eeh}) and (\ref{eep}). Thus at an electron-positron collider with centre-of-mass energy of 500 GeV, the contributions of both the radion and the Higgs in the s-channel would be negligibly small. This can be seen from the third row of Table \ref{tab1}, where the $\phi$ and h contribution is in units of $10^{-9}$ fb as compared to the total cross section shown in fb units in the second row of the table.} In Table \ref{tab1}, some total cross-section values are measured in the case of the different collision energy $\sqrt{s}$ with $P_{e^{-}} = 0.8, P_{e^{+}} = - 0.3$ \cite{bsung1, Clic1}. The possible value $\sigma_{total}$ can be about 1351.330 fb as $\sqrt{s} = 500$ GeV. With the photon and Z boson contribution in s-channel, the cross-section $\sigma_{\gamma, Z}$ is much larger than $\sigma_{\phi, h}$ with the radion and Higgs boson propagators in s-channel. \textbf{In the u, t - channel contributions, the coupling $\overline{g}_{eZ}$ would be larger than the coupling contribution in the s-channel. This can be seen from the last row of the table.} \\
\hspace*{1cm}iii) Because the terms proportional to $k_{1}^{\mu} ,k_{2}^{\nu}$ in equations (\ref{pt2}) and (\ref{pt3}) vanish \cite{raha}, the dependence on $f_{4}^{V}, f_{5}^{V}$ $(V = \gamma, Z)$ has been evaluated. The polarization coefficients are chosen as $P_{e^{-}} = 0.8, P_{e^{+}} = - 0.3$. The center-of-mass energy is 500 GeV. Contours for the pair of the parameters ($f_{4}^{\gamma}$, $f_{4}^{Z}$) are showed in Fig.\ref{Fig.3}. The total cross-section which depends on pair of the parameters ($f_{5}^{\gamma}$, $f_{5}^{Z}$) has been obtained in Fig.\ref{Fig.4}. From Fig.\ref{Fig.3} and Fig.\ref{Fig.4}, the results indicate the largest total cross-section to be about 1609.57 fb, obtained for $f_{4}^{\gamma} = 2.3 \times 10^{-3}$, $f_{4}^{Z} = -4.2 \times 10^{-3}$, $f_{5}^{\gamma, Z} = 0$.\\
\begin{table}[!htb]
\centering
\caption{\label{tab1}Some typical values for the total cross-section in the $e^{+}e^{-} \rightarrow ZZ \rightarrow l^{-}l^{+}q \overline{q}$ collisions at the ILC in the case of $P_{e^{-}} = 0.8 $, $P_{e^{+}} = -0.3 $. The parameters are chosen as $f_{4}^{\gamma} = 2.4 \times 10^{-3}$, $f_{4}^{Z} = 4.2 \times 10^{-3}$, $f_{5}^{\gamma} = 2.7 \times 10^{-3}$, $f_{5}^{Z} = 8.8 \times 10^{-3}$, $h_{1}^{\gamma} = 3.6 \times 10^{-3}$, $h_{3}^{\gamma} = 1.3 \times 10^{-3}$, $h_{1}^{Z} = 2.9 \times 10^{-3}$, $h_{3}^{Z} = 2.8 \times 10^{-3}$.}  
\begin{tabular}{|c|c|c|c|c|c|c|} 
\hline 
$\sqrt{s}$ (GeV) & 500&600&700&800&900&1000 \\ 
\hline 
$\sigma_{total} $ ($e^{+}e^{-} \rightarrow ZZ \rightarrow l^{-}l^{+}q \overline{q}$) (fb) &1351.330& 1216.980&1062.970& 896.424&722.141&545.542\\
\hline
$\sigma_{s\phi, h} $ ($e^{+}e^{-} \rightarrow ZZ \rightarrow l^{-}l^{+}q \overline{q}$) ($10^{-9}$ fb) &3.355&3.471&3.544&3.592&3.627&3.651\\
\hline
$\sigma_{s\gamma, Z} $ ($e^{+}e^{-} \rightarrow ZZ \rightarrow l^{-}l^{+}q \overline{q}$) (fb) &0.832&1.859&3.598&6.310&10.298&15.905\\
\hline
$\sigma_{u,t} $ ($e^{+}e^{-} \rightarrow ZZ \rightarrow l^{-}l^{+}q \overline{q}$) (fb) &1350.497&1215.121&1059.371&896.114&711.842&529.637\\
\hline
\end{tabular}
\end{table}
\subsection{The $e^{-}e^{+} \rightarrow \gamma Z \rightarrow \gamma l^{+} l^{-} \left(\gamma q\overline{q}\right)$ collision}
\begin{figure}[!htb] 
\begin{center}
\includegraphics[width= 7 cm,height= 5 cm]{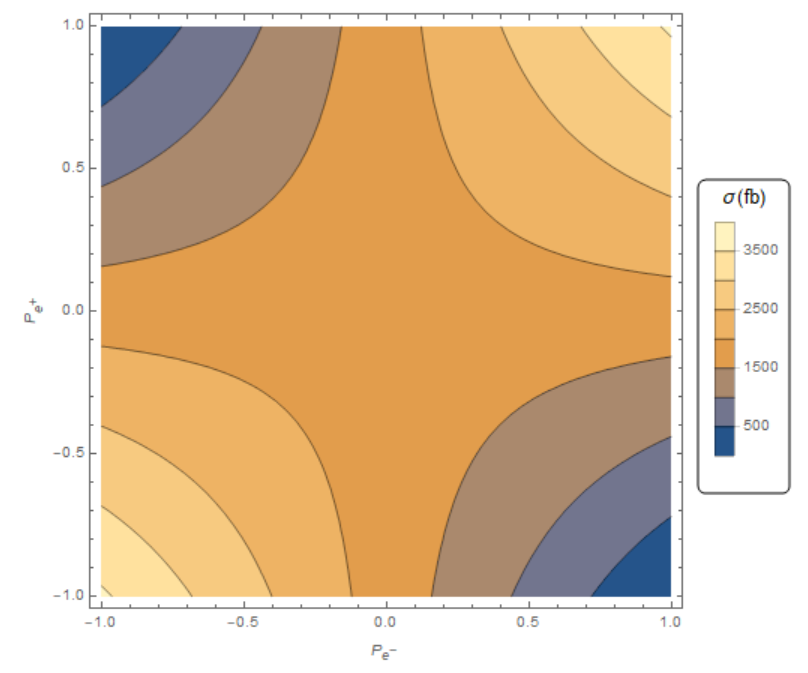}
\caption{\label{Fig.2} The total cross-section as a function of the polarization coefficients ($P_{e^{-}}, P_{e^{+}}$ ) in $e^{+}e^{-} \rightarrow ZZ \rightarrow l^{-}l^{+}q \overline{q}$ collision. The collision energy is chosen as $\sqrt{s}$ = 500 GeV. }
\end{center}
\end{figure}

\begin{figure}[!htb] 
\begin{center}
\includegraphics[width= 7 cm,height= 5 cm]{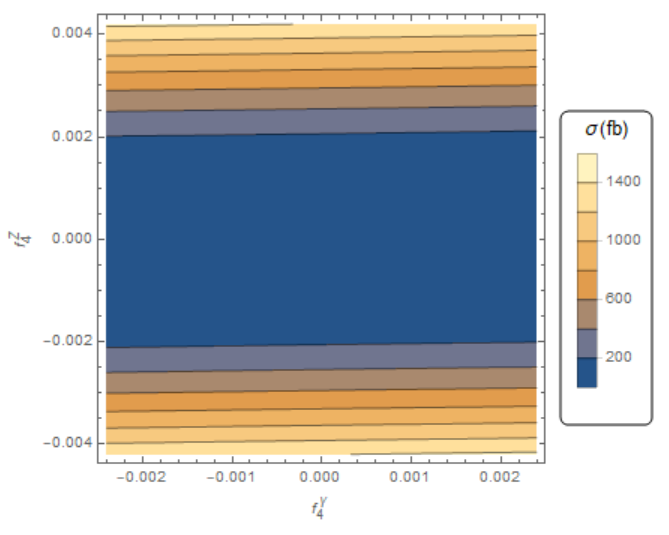}
\caption{\label{Fig.3} The total cross-section as a function of the parameters ($f_{4}^{\gamma}, f_{4}^{Z}$) in $e^{+}e^{-} \rightarrow ZZ \rightarrow l^{-}l^{+}q \overline{q}$ collision in the case of $f_{5}^{\gamma} = 2.7 \times 10^{-3}$, $f_{5}^{Z} = 8.8 \times 10^{-3}$. The parameters are taken to be $P_{e^{-}} = 0.8, P_{e^{+}} = - 0.3$, $\sqrt{s}$ = 500 GeV.}
\end{center}
\end{figure}

\begin{figure}[!htb] 
\begin{center}
\includegraphics[width= 7 cm,height= 5 cm]{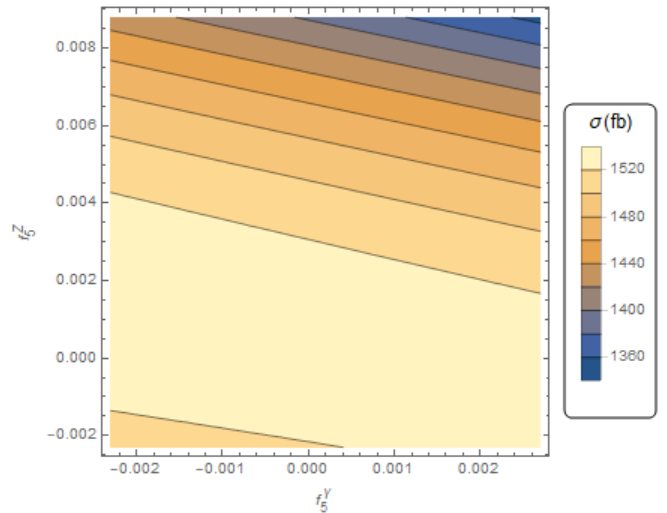}
\caption{\label{Fig.4} The total cross-section as a function of the parameters ($f_{5}^{\gamma}, f_{5}^{Z}$) in $e^{+}e^{-} \rightarrow ZZ \rightarrow l^{-}l^{+}q \overline{q}$ collision in the case of $f_{4}^{\gamma} = 2.3 \times 10^{-3}$, $f_{4}^{Z} = -4.2 \times 10^{-3}$. The parameters are chosen as $P_{e^{-}} = 0.8, P_{e^{+}} = - 0.3$, $\sqrt{s}$ = 500 GeV.}
\end{center}
\end{figure}

\begin{figure}[!htb] 
\begin{center}
\includegraphics[width= 14 cm,height= 5 cm]{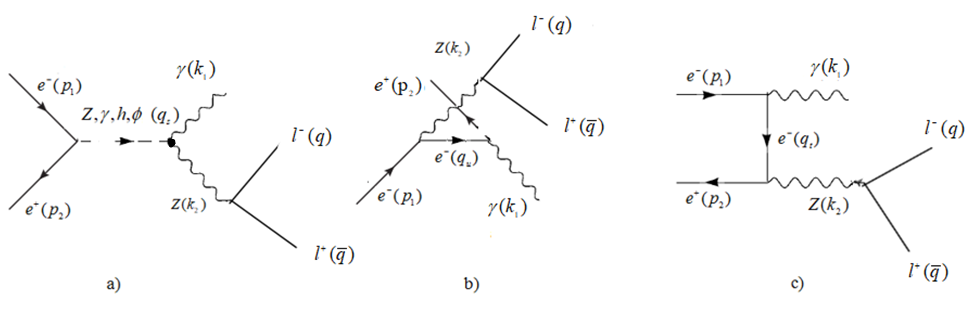}
\caption{\label{Fig.5} Feynman diagrams for $e^{+}e^{-} \rightarrow \gamma Z \rightarrow \gamma l^{-}l^{+} (\gamma q \overline{q})$ collision,representing the s, u, t-channels, respectively.}
\end{center}
\end{figure}

\begin{figure}[!htb] 
\begin{center}
    \begin{tabular}{cc}
        \includegraphics[width=5cm, height= 4cm]{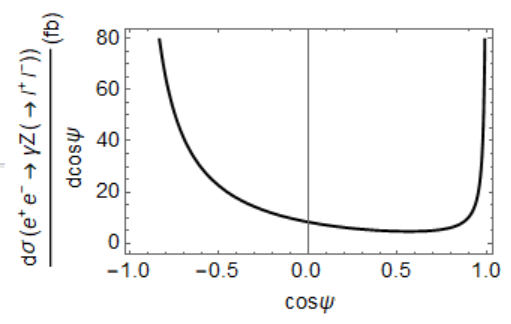} &
        \includegraphics[width=5cm, height= 4cm]{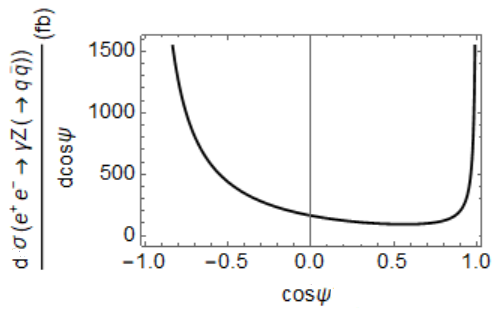} 
    \end{tabular}
    \caption{\label{Fig.6} The differential cross-section for the $e^{+}e^{-} \rightarrow \gamma Z \rightarrow \gamma l^{-}l^{+} (\gamma q \overline{q})$ collisions with respect to the $cos\psi$. The parameters are chosen as $P_{e^{-}} = 0.8, P_{e^{+}} = - 0.3$, $\sqrt{s}$ = 500 GeV.}
    \end{center}
\end{figure}

\begin{figure}[!htb] 
\begin{center}
    \begin{tabular}{cc}
        \includegraphics[width=6cm, height= 4cm]{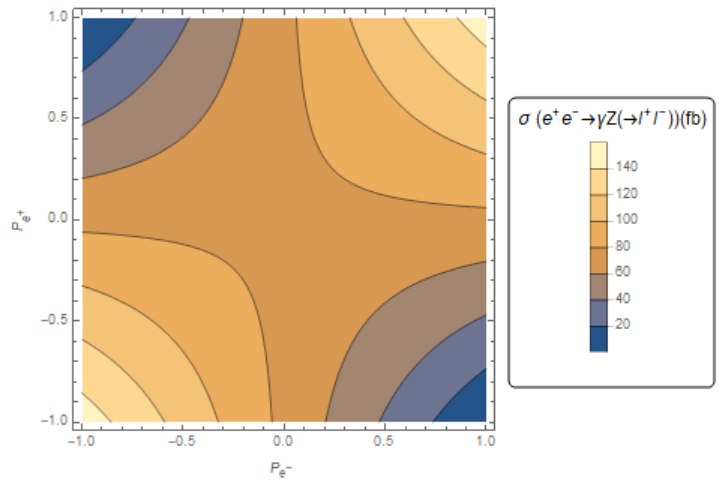} &
        \includegraphics[width=6cm, height= 4cm]{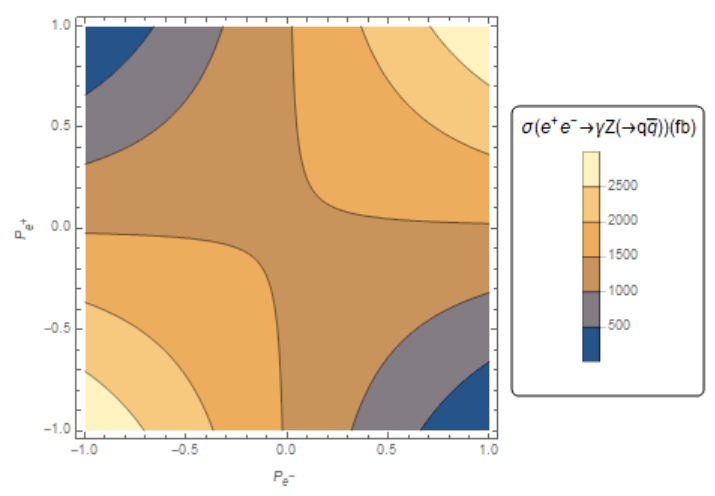} 
    \end{tabular}
    \caption{\label{Fig.7} The total cross-section as a function of the polarization coefficients ($P_{e^{-}}, P_{e^{+}}$ ) in $e^{+}e^{-} \rightarrow \gamma Z \rightarrow \gamma l^{-}l^{+} (\gamma q \overline{q})$ collisions. The collision energy is chosen as $\sqrt{s}$ = 500 GeV.}
    \end{center}
\end{figure}

\begin{figure}[!htb] 
\begin{center}
    \begin{tabular}{cc}
        \includegraphics[width=5cm, height= 4cm]{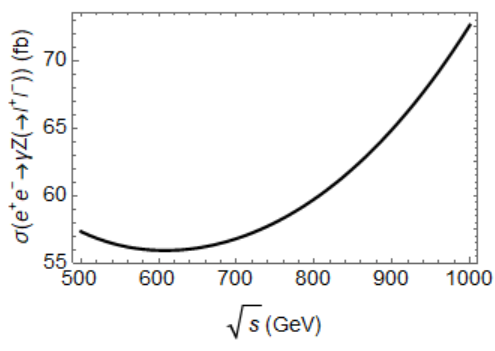} &
        \includegraphics[width=5cm, height= 4cm]{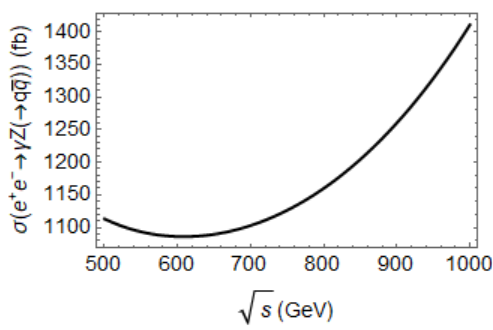} 
    \end{tabular}
    \caption{\label{Fig.8} The total cross-section for the whole process $e^{+}e^{-} \rightarrow \gamma Z \rightarrow \gamma l^{-}l^{+} (\gamma q \overline{q})$ collisions with respect to the collision energy $\sqrt{s}$. The parameters are taken to be $P_{e^{-}} = 0.8, P_{e^{+}} = - 0.3$.}
    \end{center}
\end{figure}

\begin{figure}[!htb] 
\begin{center}
\includegraphics[width= 7 cm,height= 4 cm]{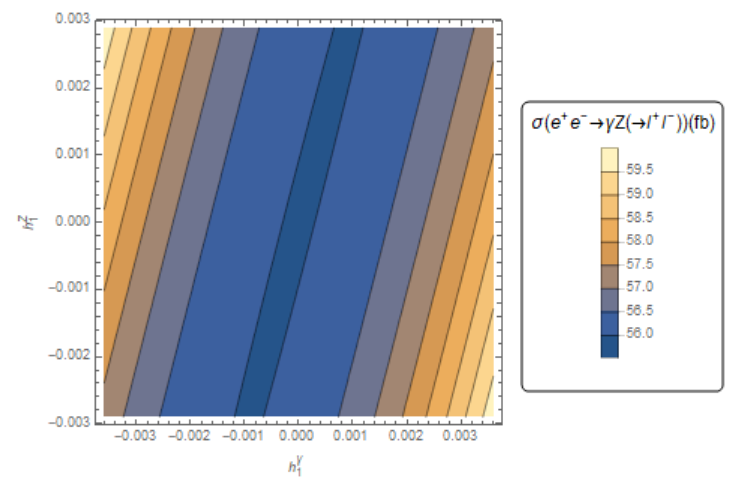}
\caption{\label{Fig.9} The total cross-section as a function of the parameters ($h_{1}^{\gamma}, h_{1}^{Z}$) in $e^{+}e^{-} \rightarrow \gamma Z \rightarrow \gamma l^{-}l^{+} $ collision in the case of $h_{3}^{\gamma} = 1.3 \times 10^{-3}$, $h_{3}^{Z} = 2.8 \times 10^{-3}$ . The parameters are fixed to $P_{e^{-}} = 0.8, P_{e^{+}} = - 0.3$, $\sqrt{s}$ = 500 GeV.}
\end{center}
\end{figure}

\begin{figure}[!htb] 
\begin{center}
\includegraphics[width= 7 cm,height= 4 cm]{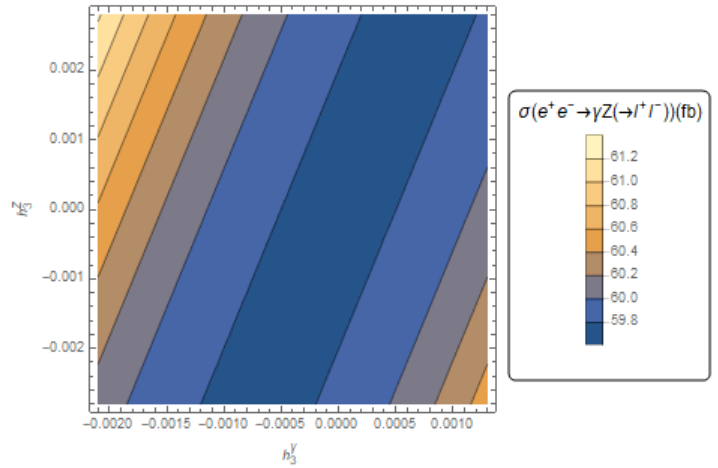}
\caption{\label{Fig.10} The total cross-section as a function of the parameters ($h_{3}^{\gamma}, h_{3}^{Z}$) in $e^{+}e^{-} \rightarrow \gamma Z \rightarrow \gamma l^{-}l^{+} $ collision in the case of $h_{1}^{\gamma} = -3.6 \times 10^{-3}$, $h_{1}^{Z} = 2.9 \times 10^{-3}$ . The parameters are fixed to $P_{e^{-}} = 0.8, P_{e^{+}} = - 0.3$, $\sqrt{s}$ = 500 GeV.}
\end{center}
\end{figure}

\hspace*{1cm}We consider the collision process in which the initial state contains electron and positron, the final state contains photon and Z boson
\begin{equation} \label{pt4}
e^{-}(p_{1}) + e^{+}(p_{2}) \    \rightarrow        \  \gamma (k_{1}) + Z (k_{2}).
\end{equation}
\hspace*{1cm}There are three Feynman diagrams contributing to reaction (\ref{pt4}), representing the s, u, t-channels exchange depicted in Fig.\ref{Fig.5}. The transition amplitude representing the s-channel is given by
\begin{equation}
M_{s} = M_{Z} + M_{\gamma} + M_{h} + M_{\phi},
\end{equation}
here 
\begin{align}
&M_{Z} = \dfrac{-\overline{g}_{eZ}}{q^{2}_{s} - m^{2}_{Z}} \varepsilon^{*}_{\mu} (k_{1}) \Gamma^{\sigma\mu\nu}_{\gamma ZZ}(q_{s} k_{1} k_{2})\varepsilon^{*}_{\nu} (k_{2})\left(\eta_{\sigma\beta} - \dfrac{q_{s\sigma}q_{s\beta}}{m^{2}_{Z}}\right)\overline{v}(p_{2}) \gamma^{\beta}\left(v_{e} - a_{e}\gamma^{5}\right) u(p_{1}),\\
&M_{\gamma} = \dfrac{-e}{q^{2}_{s}} \varepsilon^{*}_{\mu} (k_{1}) \Gamma^{\sigma\mu\nu}_{\gamma \gamma Z}(q_{s}k_{1} k_{2})\varepsilon^{*}_{\nu} (k_{2})\eta_{\sigma\beta} \overline{v}(p_{2}) \gamma^{\beta} u(p_{1}),\\
&M_{h} = \dfrac{\overline{g}_{eeh} C_{\gamma Zh}}{q^{2}_{s} - m^{2}_{h}}\varepsilon^{*}_{\mu} (k_{1}) \left[k_{1}k_{2} \eta^{\mu\nu} - k_{1}^{\nu}k_{2}^{\mu}\right]\varepsilon^{*}_{\nu} (k_{2})\overline{v}(p_{2})u(p_{1}),\\
&M_{\phi} = \dfrac{\overline{g}_{ee\phi} C_{\gamma Z \phi}}{q^{2}_{s} - m^{2}_{\phi}}\varepsilon^{*}_{\mu} (k_{1}) \left[k_{1}k_{2} \eta^{\mu\nu} - k_{1}^{\nu}k_{2}^{\mu}\right]\varepsilon^{*}_{\nu} (k_{2})\overline{v}(p_{2})u(p_{1}),
\end{align}
where
\begin{equation} \label{pt5}
\begin{aligned}
\Gamma^{\sigma\mu\nu}_{\gamma \gamma Z}&(q_{s}k_{1}k_{2}) =\\
&\dfrac{g_{e}}{m_{Z}^{2}}\Biggl[h_{1}^{\gamma}\Biggl(q_{s}^{\sigma}q_{s}^{\mu}k_{1}^{\nu}+q_{s}^{\nu}k_{1}^{\sigma}k_{1}^{\mu}-\eta^{\sigma\mu}\left(q_{s}^{2}k_{1}^{\nu}-k_{1}^{2}q_{s}^{\nu}\right)+\eta^{\sigma\nu} \left(k_{1}^{2}q_{s}^{\mu} - k_{1}q_{s}k_{1}^{\mu}\right)+\eta^{\mu\nu}\left(q_{s}^{2}k_{1}^{\sigma} - k_{1}q_{s}q_{s}^{\sigma}\right)\Biggr)\\
&- h_{3}^{\gamma}\Biggl(k_{1\rho}k_{1}^{\mu}q_{s\alpha}\varepsilon^{\sigma\nu\alpha\rho} + q_{s}^{\sigma}k_{1\alpha}q_{s\rho}\varepsilon^{\mu\nu\alpha\rho} + \left(q_{s}^{2}k_{1\alpha} - k_{1}^{2}q_{s\alpha}\right)\varepsilon^{\sigma\mu\nu\alpha}\Biggr)\Biggr],
\end{aligned}
\end{equation}

\begin{equation} \label{pt6}
\begin{aligned}
\Gamma^{\mu\nu\sigma}_{\gamma ZZ}&( k_{1}k_{2}q_{s}) = \\
&\dfrac{g_{e}}{m_{Z}^{2}}\Biggl[f_{4}^{\gamma}\Biggl(\left(q_{s}^{\nu}\eta^{\mu\sigma} + k_{2}^{\sigma}\eta^{\mu\nu}\right) k_{1}^{2} - k_{1}^{\mu} \left(k_{2}^{\sigma}k_{1}^{\nu} + q_{s}^{\nu}k_{1}^{\sigma}\right)\Biggr) + f_{5}^{\gamma}\left(k_{1}^{\mu}k_{1\rho}\varepsilon^{\nu\sigma\alpha\rho} + k_{1}^{2}\varepsilon^{\mu\nu\sigma\alpha} \right)(k_{2\alpha} - q_{s\alpha}) \\
& + h_{1}^{Z} \Biggl(q_{s}^{\mu}k_{1}^{\nu}q_{s}^{\sigma} + k_{2}^{\mu}k_{2}^{\nu}k_{1}^{\sigma} + \left(k_{2}^{2} - q_{s}^{2}\right)\left(k_{1}^{\nu} \eta^{\mu\sigma} - k_{1}^{\sigma}\eta^{\mu\nu}\right) - q_{s}^{\sigma}\eta^{\mu\nu}(k_{1}q_{s}) - k_{2}^{\nu}\eta^{\mu\sigma}(k_{1}k_{2})\Biggr)\\
& - h_{3}^{Z}\left(k_{2}^{\nu}k_{2\rho}k_{1\alpha}\varepsilon^{\mu\sigma\alpha\rho} + q_{s\rho}q_{s}^{\sigma}k_{1\alpha}\varepsilon^{\mu\nu\alpha\rho} + \left(k_{1}^{2}k_{1\alpha} - q_{s}^{2}k_{1\alpha}\right)\varepsilon^{\mu\nu\sigma\alpha}\right)\Biggr],
\end{aligned}
\end{equation}
\begin{equation}
C_{\gamma Zh} = \dfrac{\alpha}{2\pi\nu_{0}}\left[2g_{h}^{r}\left(\dfrac{b_{2}}{tan \theta_{W}} - b_{Y}tan \theta_{W}\right)-g_{h}\left(A_{F} + A_{W}\right)\right],
\end{equation}
\begin{equation}
C_{\gamma Z\phi} = \dfrac{\alpha}{2\pi\nu_{0}}\left[2g_{\phi}^{r}\left(\dfrac{b_{2}}{tan \theta_{W}} - b_{Y}tan \theta_{W}\right)-g_{\phi}\left(A_{F} + A_{W}\right)\right].
\end{equation}
\hspace*{1cm}The transition amplitude representing the u-channel can be written as
\begin{equation}
M_{u} = -i\dfrac{e\overline{g}_{eZ}}{{q^{2}_{u} - m^{2}_{e}}}\overline{v}(p_{2})\gamma^{\mu}\left(v_{e}-a_{e}\gamma^{5}\right)\varepsilon^{*}_{\mu} (k_{1})\left(\slashed{q}_{u}+m_{e}\right)\gamma^{\nu}\varepsilon^{*}_{\nu} (k_{2})u(p_{1}).
\end{equation}
\hspace*{1cm}The transition amplitude representing the t-channel is given by
\begin{equation}
M_{t} = -i\dfrac{e\overline{g}_{eZ}}{{q^{2}_{t} - m^{2}_{e}}}\overline{v}(p_{2})\gamma^{\nu}\varepsilon^{*}_{\nu} (k_{2})\left(\slashed{q}_{t}+m_{e}\right)\gamma^{\mu}\left(v_{e}-a_{e}\gamma^{5}\right) \varepsilon^{*}_{\mu} (k_{1})u(p_{1}).
\end{equation}
The total cross-section for the whole process $e^{-}e^{+} \rightarrow \gamma Z \rightarrow \gamma l^{+} l^{-} \left(\gamma q\overline{q}\right)$ can be calculated as follow \cite{raha}
\begin{equation}
\sigma = \sigma (e^{-} e^{+} \rightarrow  \gamma Z) \times Br(Z\rightarrow l^{-}l^{+} (q \overline{q})).
\end{equation}
\hspace*{1cm}We estimate the total cross-section for $\gamma l^{-}l^{+}$ production as follows \\
\hspace*{1cm}i) With the parameters $f_{4}^{V}, f_{5}^{V}, h_{1}^{V}, h_{3}^{V}$ $(V = \gamma, Z)$ as Fig.\ref{Fig.2}, $P_{e^{-}} = 0.8, P_{e^{+}} = - 0.3$, the differential cross-sections as the function of $cos\psi$ can be seen in Fig.\ref{Fig.6}. In our numerical estimation for $e^{-}e^{+} \rightarrow \gamma Z \rightarrow \gamma l^{+} l^{-}$, the lowest value of the differential cross-section can get 4.759 fb for $cos\psi$ is about 0.565. The maximum value reaches 702.233 fb when $cos\psi$ is about -0.98.\\
\hspace*{1cm}ii) The parameters $f_{4}^{V}, f_{5}^{V}, h_{1}^{V}, h_{3}^{V}$ $(V = \gamma, Z)$ are taken as Fig.\ref{Fig.2}. The Fig.\ref{Fig.7} indicates that the total cross-sections achieve the maximum value when $P_{e^{-}} = P_{e^{+}} = \pm 1$ and the minimum value when $P_{e^{-}} = 1, P_{e^{+}} = -1$ or $P_{e^{-}} = -1, P_{e^{+}} = 1$. These results are similar to the consequence of $e^{-}e^{+} \rightarrow ZZ$ collision. \textbf{The  possible value of the cross section to be 155.856 fb, obtained for $f_{4}^{\gamma, Z} = 0$, $f_{5}^{\gamma, Z} = 0$, $h_{1}^{\gamma, Z} = 0$, $h_{3}^{\gamma} = -4.2 \times 10^{-3}$, $h_{3}^{Z} = 0$, $\sqrt{s} = 500$ GeV, $P_{e^{-}} = P_{e^{+}} = \pm 1$. This value is larger than the observable theoretical value (112.40 fb) in Ref.\cite{raha} due to the radion, Higgs contributions.}\\
\hspace*{1cm}iii) With the parameters $f_{4}^{V}, f_{5}^{V}, h_{1}^{V}, h_{3}^{V}$ $(V = \gamma, Z)$ as Fig.\ref{Fig.2}, $P_{e^{-}} = 0.8, P_{e^{+}} = - 0.3$, the total cross-sections are measured in the case of the collision energy $\sqrt{s}$ in Fig.\ref{Fig.8}. The cross-sections decrease in the region $500 GeV \leq \sqrt{s} \leq 600 GeV$, then increase in the region $600 GeV \leq \sqrt{s} \leq 1000 GeV$. \\
\hspace*{1cm}iv) The polarization coefficients are chosen as $P_{e^{-}} = 0.8, P_{e^{+}} = - 0.3$. Contours for the pair of the parameters $(h_{1}^{\gamma}, h_{1}^{Z})$ is showed in Fig.\ref{Fig.9}. The maximum cross-section for  $e^{-}e^{+} \rightarrow \gamma Z \rightarrow \gamma l^{+} l^{-}$ is about 59.5 fb in the case of $h_{1}^{\gamma} = -3.6 \times 10^{-3}$, $h_{1}^{Z} = 2.9 \times 10^{-3}$ and vice versa. The total cross-section which depends on pair of the parameters $(h_{3}^{\gamma}, h_{3}^{Z})$ has been obtained in Fig.\ref{Fig.10}. The results show the largest total cross-section for $e^{-}e^{+} \rightarrow \gamma Z \rightarrow \gamma l^{+} l^{-}$ collision to be about 61.2 fb at ILC 500 GeV, obtained for $h_{1}^{\gamma} = -3.6 \times 10^{-3}$, $h_{1}^{Z} = 2.9 \times 10^{-3}$, $h_{3}^{\gamma} = -2.1 \times 10^{-3}$, $h_{3}^{Z} = 2.8 \times 10^{-3}$.\\
\hspace*{1cm}v)	Some numerical values for total cross-section in $e^{-}e^{+} \rightarrow \gamma Z \rightarrow \gamma l^{+} l^{-}$ are shown in Table \ref{tab2}. With the photon and Z boson contribution in s-channel, the cross-section $\sigma_{\gamma, Z}$ is much larger than that with radion and Higgs contribution in s-channel propagators. \textbf{Similar to $e^{-}e^{+} \rightarrow ZZ \rightarrow l^{+} l^{-} q\bar{q}$ collision, the contributions of both the radion and the Higgs in the s-channel would be negligibly small. The u, t-channels give main contribution due to the coupling $\overline{g}_{eZ}$.}
\begin{table}[!htb]
\centering
\caption{\label{tab2}Some typical values for the total cross-section in the $e^{+}e^{-} \rightarrow \gamma Z \rightarrow \gamma l^{-}l^{+}$ collisions at the ILC in the case of $P_{e^{-}} = 0.8 $, $P_{e^{+}} = -0.3 $. The parameters are chosen as $f_{4}^{\gamma} = 2.4 \times 10^{-3}$, $f_{4}^{Z} = 4.2 \times 10^{-3}$, $f_{5}^{\gamma} = 2.7 \times 10^{-3}$, $f_{5}^{Z} = 8.8 \times 10^{-3}$, $h_{1}^{\gamma} = 3.6 \times 10^{-3}$, $h_{3}^{\gamma} = 1.3 \times 10^{-3}$, $h_{1}^{Z} = 2.9 \times 10^{-3}$, $h_{3}^{Z} = 2.8 \times 10^{-3}$.} 
\begin{tabular}{|c|c|c|c|c|c|c|} 
\hline 
$\sqrt{s}$ (GeV) & 500&600&700&800&900&1000 \\ 
\hline 
$\sigma_{total} $ ($e^{+}e^{-} \rightarrow \gamma Z \rightarrow \gamma l^{-}l^{+}$) (fb) &57.333& 55.939&56.812& 59.745&64.907&72.655\\
\hline
$\sigma_{s\phi, h} $ ($e^{+}e^{-} \rightarrow \gamma Z \rightarrow \gamma l^{-}l^{+}$) ($10^{-13}$ fb) &3.303 &3.320 &3.331 &3.337 &3.342 &3.345 \\
\hline
$\sigma_{s\gamma, Z} $ ($e^{+}e^{-} \rightarrow \gamma Z \rightarrow \gamma l^{-}l^{+}$) (fb) &1.417 &3.005 &5.642 &9.707 &15.639 &23.935 \\
\hline
$\sigma_{u, t} $ ($e^{+}e^{-} \rightarrow \gamma Z \rightarrow \gamma l^{-}l^{+}$) (fb) &55.916 &52.933 & 51.169 & 50.038 & 49.268 &48.720 \\
\hline
\end{tabular} 
\end{table}

\section{Conclusion}
\hspace*{1cm} In this paper, we have studied the contribution of the couplings on the Z production at ILC in the RS model. The results show that  u, t- channels have given the main contribution in high energy \textbf{due to the electron exchanges}. With the available value of parameters $f_{4}^{V}, f_{5}^{V}$ $(V = \gamma, Z)$ , the total cross-section can be reached the observable value. The total cross-section for $e^{+}e^{-} \rightarrow ZZ \rightarrow l^{-}l^{+}q \overline{q}$ process is larger than that for $e^{+}e^{-} \rightarrow \gamma Z \rightarrow \gamma l^{-}l^{+}(\gamma q \overline{q})$  under the same conditions. This is due to $hZZ, \phi ZZ$ couplings which are much larger than $\gamma Zh, \gamma Z\phi$ vertices. \\
\hspace*{1cm} Finally, it is worth noting that the anomalous couplings of the SM gauge boson ($\gamma, Z$) \textbf{have been expected to be larger compared to the radion, Higgs contribution}. This result fits into the previous work Ref.\cite{raha}. \\

{\bf Acknowledgements}: The work is supported in part by Hanoi National University of Education under Grant No. SPHN21 – 07.\\

\newpage

\end{document}